\documentclass[prc,aps,amsfonts,amsmath,superscriptaddress,floatfix,twocolumn]{revtex4-2}
\usepackage[english]{babel}
\usepackage{graphicx,float} 

\usepackage{epsfig} 
\usepackage{rotating}
\usepackage{mwe}    
\usepackage{subfig}
\usepackage[dvipsnames]{xcolor}
\renewcommand{\d}{\rm{d}}

\newcommand{\bra}[1]{\left\langle #1 \right\vert} 			
\newcommand{\ket}[1]{\left\vert #1 \right\rangle} 			

\makeatletter
\newcommand*{\rom}[1]{\expandafter\@slowromancap\romannumeral #1@}
\makeatother

\begin{document}

\title{Patterns of spin and pseudo-spin symmetries in nuclear relativistic mean-field approaches }

\author{L. Heitz}
\affiliation{IJCLab, Universit\'e Paris-Saclay, CNRS/IN2P3, 91405 Orsay Cedex, France}
\affiliation{IRFU, CEA, Universit\'e Paris-Saclay, 91191 Gif-sur-Yvette, France}
\author{J.-P. Ebran}
\affiliation{CEA,DAM,DIF, F-91297 Arpajon, France}
\affiliation{Universit\'e Paris-Saclay, CEA, Laboratoire Mati\`ere en Conditions Extr\^emes, 91680, Bruy\`eres-le-Ch\^atel, France}
\author{E. Khan}
\affiliation{IJCLab, Universit\'e Paris-Saclay, CNRS/IN2P3, 91405 Orsay Cedex, France}
\affiliation{Institut Universitaire de France (IUF)}

\begin{abstract} 
The behavior of spin doublets is known to play a major role in nuclear structure and shell effects. Pseudo-spin doublets are also known to impact the 
single-particle spectrum. The covariant framework, having these two effects encoded in its approach, is an excellent tool to understand the main mechanism
driving theses spin and pseudo-spin symmetries and their breaking. A perturbative expansion of the degeneracy raising related to spin and pseudo-spin 
effects is proposed, up to second order. It allows to understand the main behavior of spin and pseudo-spin energy doublets, such as their A dependence, as well as their common footing and differences. In the case of the spin symmetry, only the 
lower component of the Dirac bi-spinor is involved, whereas in the case of the pseudo-spin one, both the upper and lower components are involved. Their interplay with the covariant potentials is also analyzed.
\end{abstract}
 


\date{\today}

\maketitle
\section{Introduction}

The concepts of shell structure and single particle levels provide an intuitive understanding of nuclear structure. For instance, magic numbers are associated to significant gaps in single-particle energies near the Fermi level. Even if such single particle states are not proper observables  and are model-dependent \cite{baranger1970,duguet2012,duguet2015,soma2024}, they nonetheless offer  a valuable  perspective on nuclear structure. In particular, shell structure is shaped by the realization or breaking of underlying symmetries, which either split single states apart or render them degenerate. 

The manner in which spin symmetry is realized is known to play a pivotal role in accounting for magic numbers \cite{mayer_closed_1949,bm}, motivating the inclusion of a pronounced spin-orbit force in nuclear mean-field. Pseudo-spin symmetry (PSS) \cite{hecht1969,arima1969}, another spin-related symmetry, was initially proposed   as the   symmetry underlying  the observed quasi-degeneracy between  $(n,\ell, j = \ell + 1/2)$ and $(n-1, \ell +2 , j' = \ell +3/2)$ levels. Its origin has been traced back, within a relativistic framework, to the smallness of the confining potential $\Sigma$ \cite{ginocchio1997}. PSS has  been shown to underpin various properties of atomic nuclei, including  superdeformed configurations \cite{dudek1987,bahri1992}, identical bands \cite{nazarewicz1990,nazarewicz1990a,zeng1991}, magnetic moments and transitions \cite{vonneumann-cosel2000,troltenier1994,ginocchio1999}, shape coexistence \cite{delafosse2018} or charge radii kinks \cite{ebran2016}. Recent studies \cite{heitz2024}, suggest that PSS plays a significant role in shaping shell structure far from stability. For a discussion on the current understanding of PSS, see \cite{liang2015}. It is therefore crucial to capture how PSS is broken, or accidentally restored, across the nuclear chart.

In order to understand how these symmetries are implemented in relativistic mean-field approaches, we derive both spin-orbit (SO) and pseudo-spin-orbit (PSO) gaps within a common perturbative approach. It allows us to derive explicit approximate formulae for these gaps,  which serve to highlight the generic mechanisms governing symmetry breaking and restoration, as well as deviations from these patterns.

Section II introduces the perturbative framework and outlines the computation of energy gaps at first and second orders. Section III examines spin symmetry breaking within this framework, offering a nuanced reinterpretation of generic results. Finally, section IV explores the consequences of pseudo-spin symmetry breaking. 

\section{Perturbative framework}

Within the relativistic mean field framework, the approximate realization of  PSS has been associated with the small magnitude of the central potential $\Sigma$ \cite{meng1998} , defined as the sum of vector $V$ and scalar $S$ self-consistent potentials $\Sigma \equiv V+ S$. On the other hand, their difference $\Delta \equiv V-S$ is known to underpin spin symmetry breaking \cite{ebran2016,ebran2016a}. Nevertheless, the precise mechanism responsible for breaking PSS remains elusive. The challenge can be attributed  to the unphysical nature of exact PSS: if PSS were to hold exactly, nucleons would be unbound \cite{ginocchio1997}. Consequently, the unphysical nature of exact PSS prevents the use a PSS-presevring system as a reference state in perturbative calculations. However, it has been shown that a harmonic oscillator state provides a well-defined reference for a perturbative computation of PSS splittings \cite{liang2011}. 
The present section introduces the perturbative framework and propose a new derivation of the single-particle splitting to first and second order in perturbation theory.

\subsection{Reference state}

In Ref. \cite{liang2011}, it was suggested that, given self-consistent $\Sigma$ and $\Delta$ potentials, the spin-symmetric relativistic harmonic oscillator (SS-RHO) constitutes a genuine reference state for a perturbative treatment of PSS breaking. The associated reference potential $\Sigma = \Sigma_{\textrm{HO}}$ is harmonic, whereas $\Delta = \Delta_0$ is constant, to ensure spin-symmetry \cite{ebran2016,ebran2016a}. While this study confirmed that the perturbative framework was valid by demonstrating that perturbation parameters remained well below unity, this approach was not used to derive explicit results. In contrast, the present work elaborates on this framework by deriving closed-form expressions for PSS splittings and analyzing their contributions, offering new insights into the mechanisms underlying symmetry breaking and restoration. It is then possible to retrieve PSS gaps from self-consistent $\Sigma$ and $\Delta$ potentials, by acting perturbatively with these potentials, on top of the SS-RHO reference state. 

In the following, spherical symmetry is assumed. The reference Dirac Hamiltonian, and associated reference wave-functions are  :

\begin{equation}
H_0 = \begin{pmatrix}
\Sigma_{\textrm{HO}} +M & \frac{\d}{\d r} - \frac{\kappa}{r} \\
\frac{\d}{\d r} + \frac{\kappa}{r}  & M - \Delta_\textrm{HO}
\end{pmatrix}
,
\Psi_{n,\ell,j}(r) = \begin{pmatrix}
g_{n,\ell}(r) \\
if_{n,\ell,j}(r)
\end{pmatrix},
\label{Eq:SSRHO}
\end{equation}

\noindent where $\kappa = (\ell-j)(2j+1)$, $\Sigma_\textrm{HO} = c_0 + c_2 r^2$ and $\Delta_\textrm{HO} = d_0$ constant.  The precise expression of eigenenergies and eigenvectors can be found in \cite{ginocchio2004}. $g$ ($f$) stands for the upper (lower) component.

We mention here two properties that will prove useful throughout the remainder of the present article. i) In a non-relativistic reduction, the Schrödinger wave-function is identified  with the upper component $g$, after proper normalisation \cite{reinhard1989}, $\Sigma$ with the central potential and the spin-orbit potential is related to $\Delta$ \cite{ebran2016,ebran2016a}: 

\begin{equation}
    V_{\textrm{ls}}\sim \frac{1}{r}\frac{\d \Delta}{\d r},
    \label{Eq:Vls}
\end{equation}

ii)  Only $f$ depends on quantum number $j$, namely on spin.

\subsection{Perturbative expansion}

Given any  $\Sigma$ and $\Delta$ potentials, it is possible to act perturbatively on top of the HO reference state of the previous section, through the operator $W$ :

\begin{equation}
W = \begin{pmatrix}
\Sigma - \Sigma_{\textrm{HO}} & 0 \\
0 & d_0 - \Delta
\end{pmatrix}.
\end{equation}

\noindent Here, we chose $\Sigma,\Delta$ self-consistently determined from Relativistic Hartree Bogoliubov (RHB) approach \cite{niksic2014}, using the DD-MEV parametrisation \cite{mercier2023}. This interaction has been derived to take into account single-particle data in its design fit.

\begin{figure}[tb]
\scalebox{0.3}{\includegraphics{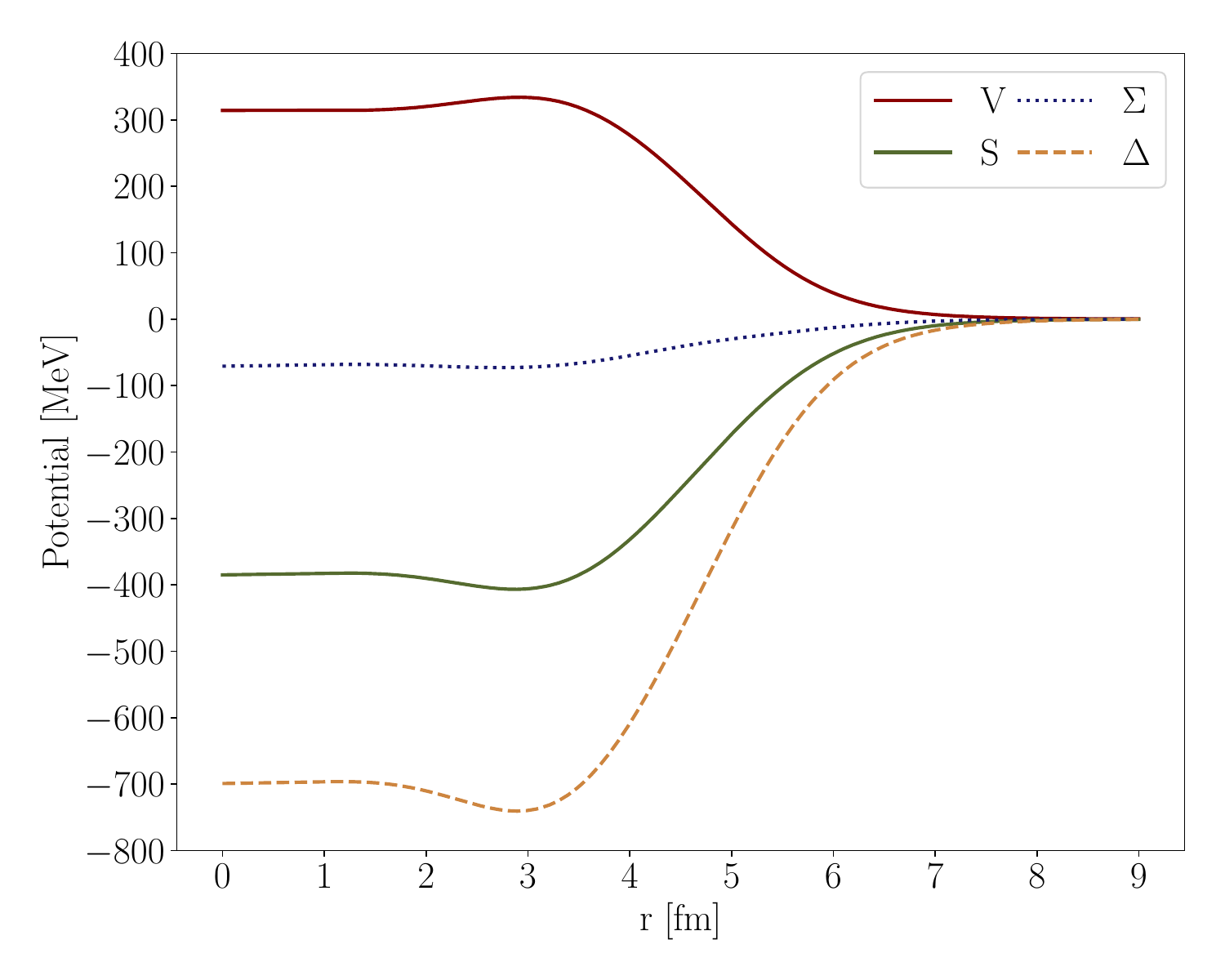}}
 \caption{Self-consistent determined neutron potentials in $^{78}$Ni, at the Relativistic Hartree Bogoliubov level. $\Sigma$ is defined as $ S+ V$, whereas $\Delta = S-V$ }
 \label{fig:VS}
\end{figure}

We use this perturbative framework to study PSS and SS breaking with self-consistent potentials. In the following, subscripts $a,b$ refer to the symmetry partners and superscripts  $^{(1,2)}$ refer to the order of the perturbation. Here, we derive the energy splitting between any two orbitals a,b at first order using perturbation theory:

\begin{eqnarray}
\begin{split}
\Delta E^{(1)} = \int \d r \; r^2 \left[ (g_a^2 - g_b^2)\left( \Sigma(r) - \Sigma_{\textrm{HO}}\right)  \right.+ \\
\left. (f_a^2 - f_b^2)\left( d_0 - \Delta(r)\right) \right]
\end{split}.
\label{Eq:PT1}
\end{eqnarray}

\noindent In order to check to the relevance of this perturbative approach, we also derive the energy gap at second order, $\Delta E^{(2)} = \delta E_a ^{(2)} - \delta E_b^{(2)}$, with

\begin{eqnarray}
\begin{split}
\delta E_c^{(2)} = \sum_{i'}\frac{1}{E_c^{(0)} - E_{i'}^{(0)}} \left[\int \d r \; r^2 \left\{ g_c g_{i'}\left( \Sigma - \Sigma_{\textrm{HO}}\right)  \right.+ \right. \\
\left.\phantom{\int}\left. f_c f_{i'}\left( d_0 - \Delta\right) \right\} \right]^2
\end{split}
\label{Eq:PT2}
\end{eqnarray}

\noindent Here $g_i$ and $f_i$ always refer to the wavefunctions at zeroth-order, namely the eigenfunctions of SS-RHO. Due to spherical symmetry, the sum runs over radial quantum number $n'$, with fixed $\ell'=\ell_c,j'=j_c$ .

\subsection{Perturbation parameter and optimisation}

There are three free parameters in the HO reference state (see Eq. (\ref{Eq:SSRHO})): $d_0,c_0$ and $c_2$. We follow \cite{liang2011} and determine them by minimizing a perturbation parameter W$_m$ which, for a given state $m$, reads:

\begin{equation}
W_m = \underset{n_p, \ell_p = \ell_m, j_p = j_m}{\textrm{max}} \left| \frac{\bra{\Psi_m} W \ket{\Psi_p}}{E_m^{(0)}-E_p^{(0)}}\right|.
\end{equation}

\noindent This definition is conservative, as it allows to provide an upper bound in absolute value for any term entering the perturbative calculation.  For the perturbative approach to be justified, the condition $W_m \ll 1$ has to be satisfied. To ensure such a criterion, the reference state parameters $c_0, c_2$ and $d_0$ are chosen as to minimize $W_m$. 

At first, for a given state and potentials $\Sigma,\Delta$, all three parameters were optimised numerically but it turned out that $c_0$ and $d_0$ converged to  $c_0 = -73$ MeV, $d_0 = -588$ MeV, within a few MeV, regardless of the state at stake. These values are in agreement with those of ref. \cite{liang2011} and are typical depths of relativistic potentials. Therefore only $c_2$ (or equivalently the harmonic oscillator frequency $\omega$) is actually optimised for each calculation. In all calculations performed, the condition $W_m<1$ has been checked. In reference \cite{liang2011}, this condition was only checked for PSS partners. We found  $W_m$ to be $\sim 10^{-2}$ for SS partners, thus justifying the perturbative approach for SS splitting with a SS-RHO reference state.

\subsection{Benchmark on the N=50 case}

\begin{figure}[tb]
\scalebox{0.35}{\includegraphics{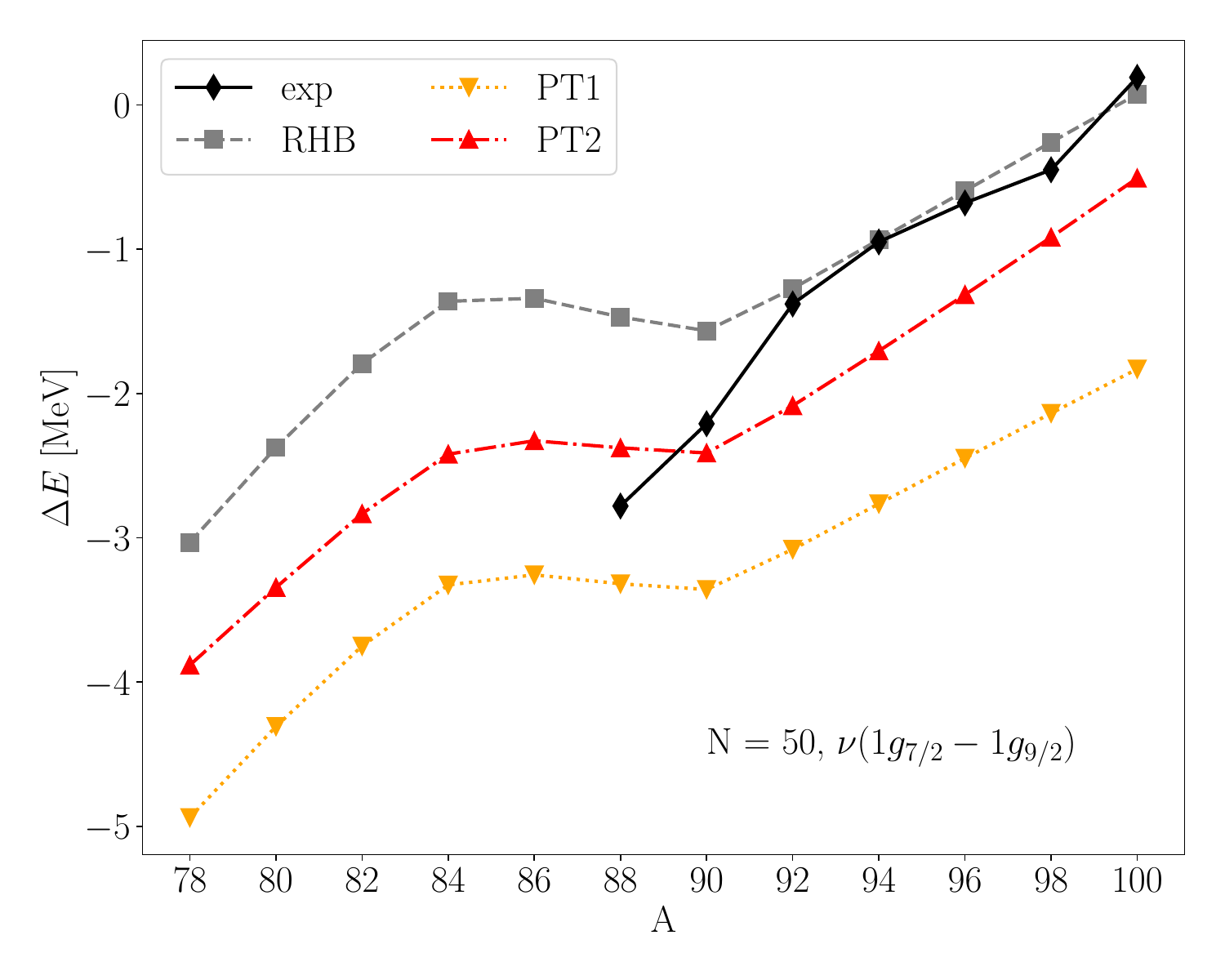}}
 \caption{ Neutron 1$\tilde{f}$ PSS gap in N = 50 isotonic chain. Grey squares correspond the RHB calculation output, red up triangles to first order in perturbation (Eq. \ref{Eq:PT1}) with RHB $\Sigma$ and $\Delta$ potentials, and orange down triangles to second order in perturbation (Eq. \ref{Eq:PT2}).}
 \label{fig:N50_PSS}
\end{figure}

In order to assess the relevance of the present perturbative approach, the N=50 case is considered. 
Neutron levels are filled up to $1g_{9/2}$. The levels right above are $1g_{7/2}$ and $2d_{5/2}$. Consequently, the $N = 50$ gap is driven by both the breaking of SS in $1g$ states and PSS in $1\tilde{f}$ states. More precisely, we consider the evolution of the $\nu(2d_{5/2}-1g_{7/2})$ PSS splitting along the N = 50 isotonic chain. RHB calculations, as well as results obtained in first (Eq. \ref{Eq:PT1}) and second (Eq. \ref{Eq:PT2}) order perturbation theory are displayed in figure \ref{fig:N50_PSS}. RHB calculations closely align with experimental values. Moreover, while there is an offset between RHB and perturbation theory results, the overall trend is still captured by perturbation, even at first order. 

The effect of the second order is a mere global shift towards the
RHB values. Therefore, the investigation of the first order perturbation approach shall already enlighten the understanding of pseudo-spin behavior, especially its A dependence. In the following, we will remain at this order to investigate both the spin and pseudo-spin symmetry breaking.

\section{Spin Symmetry breaking}

We  briefly review and recover within the present framework (first order perturbation) the mechanism of SS breaking. In this section only, partners we consider partners as $a \equiv \downarrow  \equiv (n, \ell, j = \ell - 1/2)$, $b \equiv \uparrow \equiv (n,\ell,j=\ell+1/2)$.

\subsection{General scaling of SS breaking}

In non-relativistic approaches the scaling of SS breaking with  mass number $A$ can be derived by approximating the spin-orbit potential $V_{ls}$ by a $\delta$ distribution and using a phenomenological scaling for the wavefunctions \cite{bohr1998}. In a relativistic framework, this scaling law is naturally recovered by the the reduction of the Dirac equation of $\Psi$ to a Schrödinger-like equation for the upper component $g$ \cite{ebran2016,ebran2016a}. Such an analysis leads to a typical $A^{-2/3}$ scaling for SS breaking, which is well verified experimentally \cite{mairle1993}. This key relation is encoded in
the analytical form of the spin-orbit potential, as given by Eq. (\ref{Eq:Vls}). This expression yields two $1/R \sim A^{-1/3}$ factors: one for $1/r$ and another for $\frac{\d \Delta}{\d r}$, resulting in a $A^{-2/3}$ scaling \cite{ebran2016}.

However, we are interested in recovering this $A^{-2/3}$ scaling from a covariant
analysis, i.e. without the non-relativistic reduction. This shall provide a deeper understanding for the spin symmetry breaking and prepare the 
analysis for the pseudo-spin one. On this purpose, we use the first order perturbative approach described in the previous section.

\subsection{SS breaking from SS RHO }

\subsubsection{General scaling}

It should be first reminded that at zeroth order, $\Delta E^{(0)} =0$, as the HO reference state is spin-symmetric. Let us now derive the energy gap at first order. Using the independence of the upper component g from the quantum number j (see section II.A), Eq. (\ref{Eq:PT1}) reduces to :

\begin{equation}
\Delta E^{(1)}_{\textrm{SS}} \equiv \Delta E^{(1)} = \int \d r \; r^2 (f_\downarrow^2 - f_\uparrow^2)\left( d_0 - \Delta(r)\right)
\end{equation}

\noindent which yields, after integration by parts :

\begin{equation}
\Delta E^{(1)} = \int \d r \; (F_\downarrow - F_\uparrow) \frac{\d \Delta}{\d r},
\label{Eq:PT1SS}
\end{equation}

\noindent where $F_i(r) \equiv \int _0^r \d r' \; r'^2 f_i(r')^2$ is the integrated lower-component density. Since $\Delta'(r)$ is sharply peaked at $r \simeq R$, it can be well approximated by a $\delta$ distribution \cite{ebran2016,bohr1998}, leading to:

\begin{equation}
\label{eq:dESS}
\Delta E^{(1)} \simeq \left(F_\downarrow(R) - F_\uparrow(R)\right) \Delta_0,
\end{equation}

\noindent where $\Delta_0 \sim \Delta(r=0)$ is the typical depth of $\Delta$. 

\begin{figure}[tb]
\scalebox{0.35}{\includegraphics{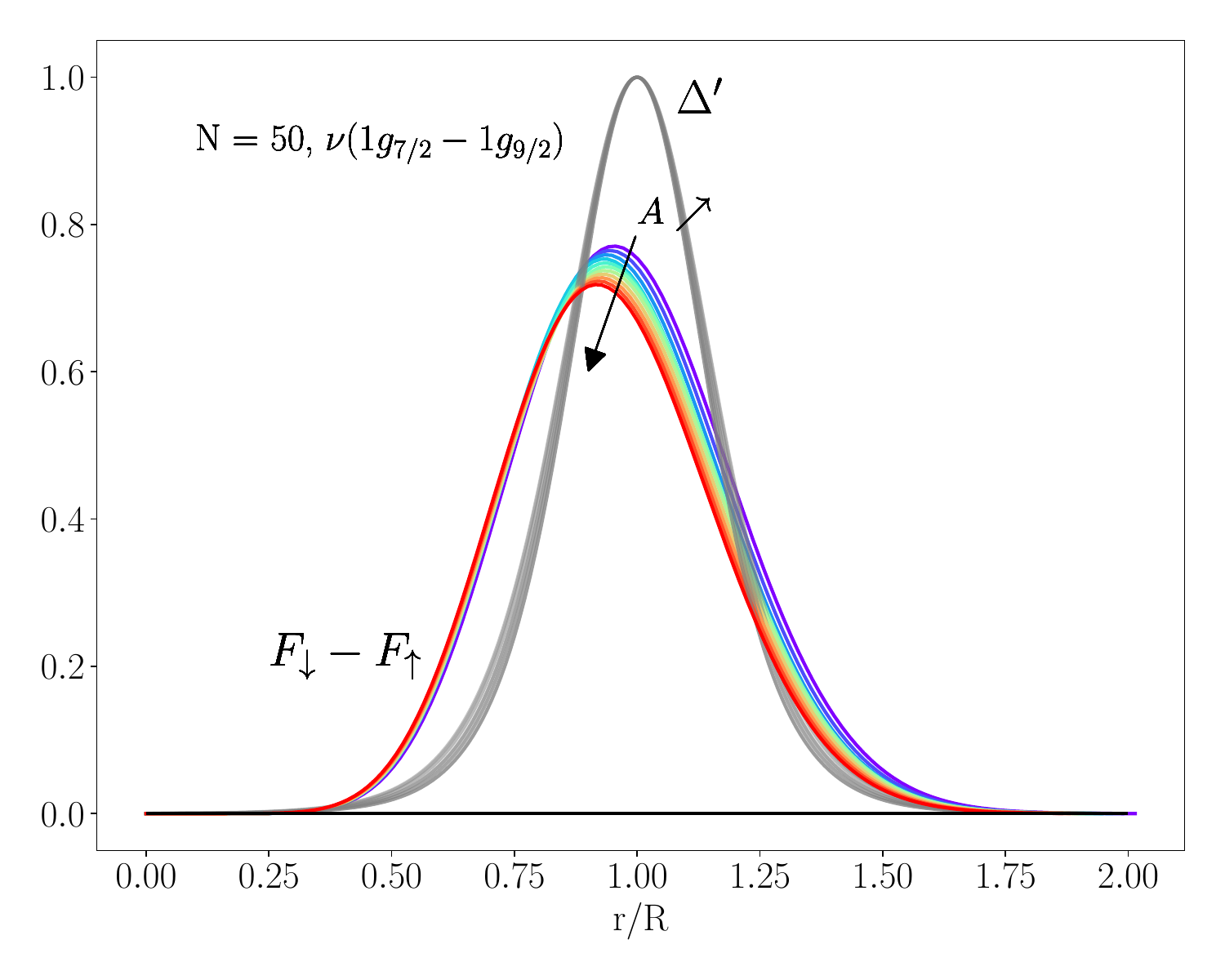}}
 \caption{Evolution along $N=50$ of $F_\downarrow- F_\uparrow$ (Eq. (\ref{eq:dESS})). $F_i =\int_0^r\d r' r'^2f(r')^2 $ is the integral of the lower component density. $\Delta'$ is the spatial derivative of $\Delta$.}
 \label{fig:F1F2}
\end{figure}

To further analyse how Eq. (\ref{eq:dESS}) depends on $A$, let us assume that $\Delta_0$ is approximately constant in an isotopic or isotonic chain. This last assumption has been checked  numerically with Relativistic Hartree Bogoliubov (RHB) potentials \cite{niksic2014} up to $20 \%$ variations. Therefore, the main variations of $\Delta E^{(1)}$ are due to the evolution of $F_\downarrow - F_\uparrow$. Indeed, the two integrands, $F_\downarrow-F_\uparrow$ and $\Delta'$, are displayed on Fig. \ref{fig:F1F2} for different values of $R$, or equivalently of $A$. As A increases, the $\Delta E^{(1)}_{\textrm{SS}}$ gap is found to decrease as $A^{-2/3}$, from two main effects: i) The norm of $F_{\uparrow\downarrow}$ decreases as $A^{-1/3}$ and ii) The overlap between $\Delta'$ and $\left(F_\downarrow(R) - F_\uparrow(R)\right)$ also decreases as $A^{-1/3}$. 

In order to explain i) and ii), let us recall  Eq. (44) from  \cite{ginocchio2004}, using the present notation, with $N$ labeling the state with  eigen-energy $E_m$  for the SS-RHO hamiltonian :

\begin{equation}
    \frac{\int\d \,r r^2   f_N^2}{\int \d r \, r^2 g_N^2} = \frac{E_N -M - c_0}{2\left(E_N + M -d_0\right)} 
\end{equation}

In the limit $\omega \ll M$, we also use equation (24) of \cite{ginocchio2004} to get $E_N -M -c_0 \simeq (N + \frac{3}{2})\omega $. Using the normalization of the wavefunction, in this limit we also have  $\int\d r \, r^2 g^2(r) \simeq 1$. We now prove i) :

\begin{equation}
\label{Eq:normf}
F_N(R)\leq  \int_0^\infty \d r \; r^2 f_N^2(r) \propto \frac{\omega}{\widetilde{M}} \left(N + \frac{3}{2}\right) \propto{\omega} \sim A^{-1/3},
\end{equation}

\noindent with $\widetilde{M} = M + (c_0-d_0)/2$ and $\omega$ the harmonic oscillator frequency that generically scales as $A^{-1/3}$. The origin of this decrease can also be qualitatively explained. As $A$ increases,the binding potential extends over a larger region, confining nucleons in a broader region $L$. Using Heisenberg relation $\Delta p \simeq \hbar /L$, this means that the associated kinetic energy $\sim (\Delta p)^2 /2m \sim \hbar^2/(2mL^2)$ is smaller. As the kinetic energy reduces, the contribution of the lower component f reduces, as it is of relativistic nature.

Effect ii) can be understood geometrically from Eq. (\ref{Eq:PT1SS}). As $A$ increases, the overlap between $\Delta'$ and $F_\downarrow-F_\uparrow$ decreases, as their maxima get shifted with respect to one another. To numerically quantify this effect, we kept the norm of  $F$  constant—eliminating the influence of effect i)—and then varied  A . This resulted in a characteristic  $A^{-1/3}$  scaling.

To sum up, the typical scaling $A^{-2/3}$ can be split in the product of two $A^{-1/3}$ factors: one coming from the weakening of the relativistic nature of system, another one from the decreasing overlap between $\Delta'$ and $F$.

Regarding the dependence on quantum numbers, the SS splitting is known to scale as $(2\ell+1)$ \cite{bm}. This can be understood within the present approach with the use of Eq. (\ref{eq:dESS}), as the norm of $F$ is linear in $\ell$. Moreover, the usual $1/n$ scaling of SS splitting can be traced back to the radial structure of $F$: as $n$ increases, the overlap with $\Delta'$ decreases because of the nodal structure of $F$, as the positive and negative contributions partially compensate.

\subsubsection{Typical trends and deviations}

\begin{figure}[tb]
\scalebox{0.35}{\includegraphics{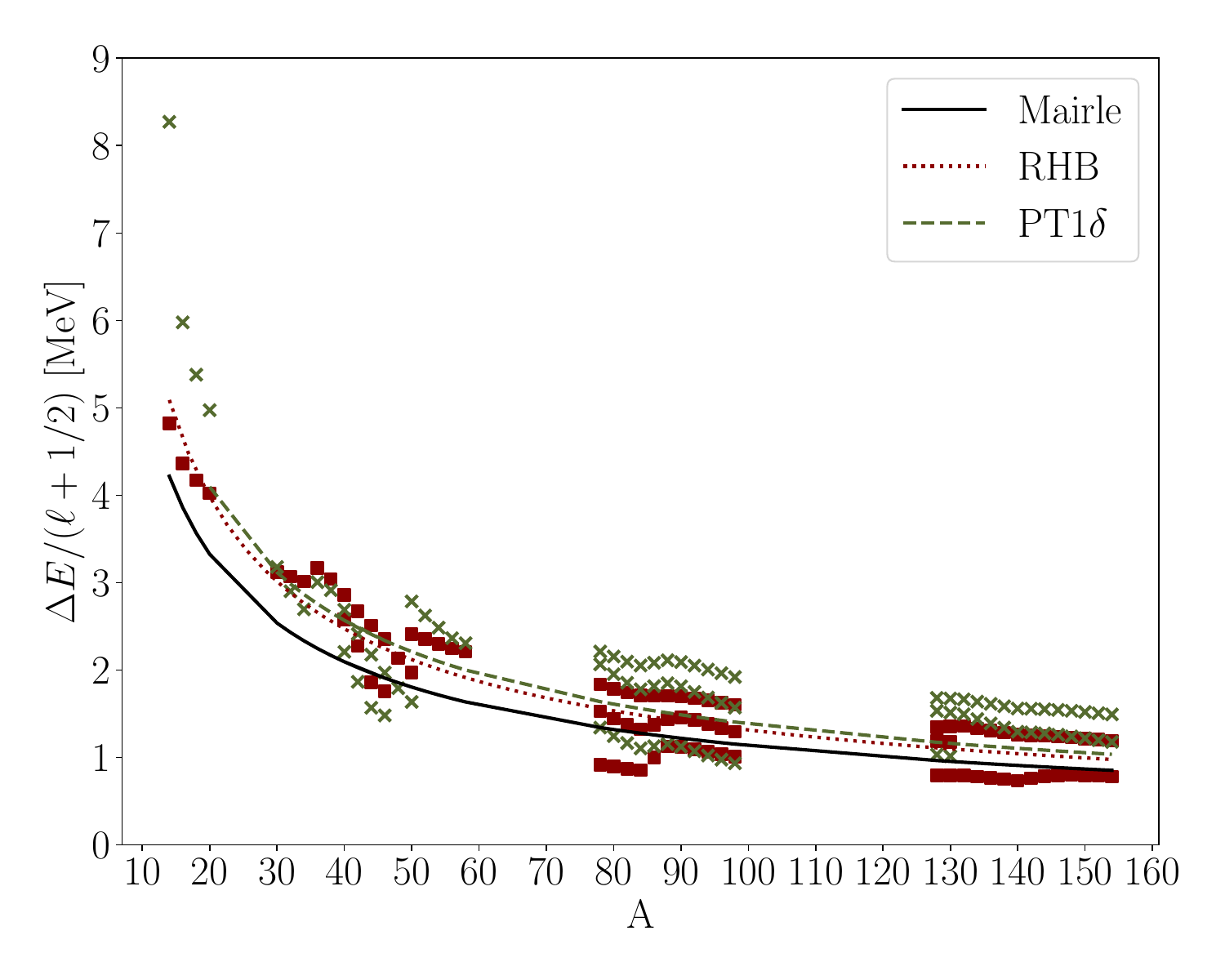}}
 \caption{Perturbation at first order for spin symmetry gap, in the $n = 1$, neutron case. Red squares correspond to RHB calculations, green crosses to Eq. (\ref{eq:dESS}). Red and green dashed lines correspond to a power-law fit $y = p_1 A^{-p_2}$ whereas black dashed lines correspond to \cite{mairle1993}. } 
 \label{fig:bulk_spin}
\end{figure}

Fig. \ref{fig:bulk_spin} shows the experimental trend of SS splitting \cite{mairle1993}, the output of RHB calculations as well as those obtained with Eq. (\ref{Eq:PT1SS}). The trend of SS splitting has been probed experimentally \cite{mairle1993} and leads to $\Delta E \sim 24.5 (\ell +1/2) A^{-2/3}$. This bulk law has been tested for $n=1$ for several $\ell$. Here, Eq. (\ref{eq:dESS}) has been applied, with a potential depth $\Delta_0$ fitted with a Woods-Saxon potential to the actual RHB potential.

Although the precise values are not captured, the theoretical trend is already reproduced at first order, and is rather consistent with the experimental one, justifying the approximations made.

It should be noted that the tensor force may also locally impact the spin-orbit splitting, see \cite{otsuka2020} and references therein. Depending on the relative proton/neutron spins, the tensor effect may either enhance or mitigate the splitting. This can generate a deviation to the $A^{-2/3}$ trend which has been observed experimentally. Another example is provided by bubble nuclei, that have an internal density depletion. This implies that replacing $\Delta'$ with a peak function at $R$ is no longer valid and the internal region of $F_\downarrow - F_\uparrow$ is probed in this case.

\section{Pseudo-Spin Symmetry breaking}

In the previous section, well-known SS splitting results have been recovered and reinterpreted within the present framework. Within the same lines, we shall turn to PSS splitting. From now on, we use $a \equiv (n, \ell, j = \ell + 1/2)$ and $b \equiv (n-1,\ell+2,j'=\ell+3/2)$. 

\subsection{Phenomenological motivations}

The large breaking of SS is known to give rise, for instance to  magic gap in stable nuclei. On the other hand, PSS is approximately realized in nuclei, which implies that PSS partners usually lie close to each other. This leads to phenomena such as identical bands \cite{nazarewicz1990,nazarewicz1990a,zeng1991}. Yet, precisely understanding how this symmetry is either restored or broken along an isotopic or isotonic chain is of major interest, as it provides key insights of shell evolution far from stability \cite{heitz2024}.

\subsection{Preliminary analysis}

The PSS breaking being more intricate, we start with a preliminary analysis of the energy splitting. Unlike the SS case, the integral over $g$ does not vanish here: the full expression in Eq. (\ref{Eq:PT1}) has to be used.

However, let us first note that the integral over $\Sigma_{\textrm{HO}}$ vanishes identically, as $g_a$ and $g_b$ are associated to the same HO eigenvalue at zeroth order: $N_b = 2\left((n-1) -1 \right) + (\ell+1) = 2(n-1) + \ell = N_a$. At odds with SS splitting, see Eq. (\ref{eq:dESS}), both $\Sigma$,$\Delta$ and both $g$,$f$ are involved in the PSS breaking, and their contributions to the PSS splitting are expected to be of the same order of magnitude. The PSS splitting at first order in perturbation theory is obtained along a similar derivation than for the SS splitting, by using integration by parts:

\begin{equation}
\begin{aligned}
\Delta E_{\textrm{PSS}}^{(1)} \equiv \Delta E^{(1)} & = \int \d r \; (G_a-G_b) \frac{\d \Sigma}{\d r} + \int  \d r \; (F_a-F_b) \frac{\d \Delta}{\d r} \\
& \equiv \Delta E_{g} +\Delta E_{f}
\end{aligned}
\label{Eq:PT1PSS}
\end{equation}

\noindent where $F$ has the same expression as in (\ref{eq:dESS}) and $G_i \equiv \int_0^r \d r' r'^2 g_i(r')^2$ is the upper-component counterpart. From now on, the splitting due to  $\Sigma$ and g ($\Delta$ and f) is coined $\Delta E_g$ ($\Delta E_f$). Both contributions are  now  analyzed in the specific case of the $\nu(2d_{5/2},1g_{7/2})\equiv(a,b)$ splitting in the experimentally relevant N = 50 isotonic chain, as stated previously.  

\subsection{General scaling with $A$}

In order to highlight the mechanisms of PSS splittings, self-consistent $\Delta$ and $\Sigma$ potentials have been fitted to Woods-Saxon potentials. This allows us to disentangle the different contributions to the PSS splitting and thus provide broad keys to understanding.  

\begin{figure}[tb]
\scalebox{0.35}{\includegraphics{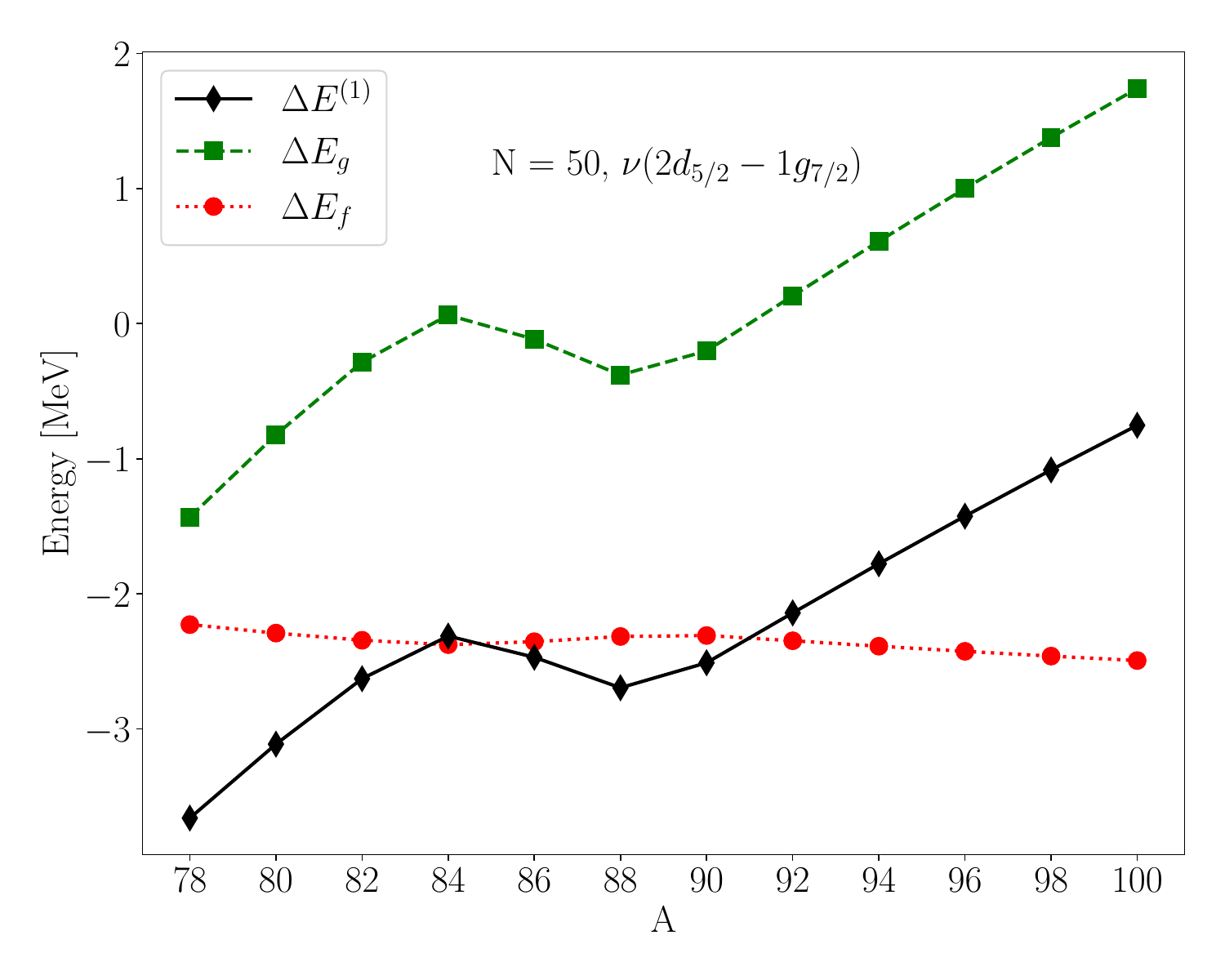}}
 \caption{ Contribution to the PSS gap of the upper component $g$ (green squares) and of the lower component $f$ (red dots).}
 \label{fig:N50_EgEf}
\end{figure}

Both $\Delta E_g$ and $\Delta E_f$ contributions are depicted on figure \ref{fig:N50_EgEf}. $\Delta E_g$ is found to scale linearly with mass number $A$ for various A ranges, being usually negative for small $A$. On the other hand, $\Delta E_f$ is generically found to be roughly constant with mass number $A$. We now turn to the analysis of each contribution.

\subsubsection{$\Delta E_g$}

\begin{figure}[tb]
\scalebox{0.35}{\includegraphics{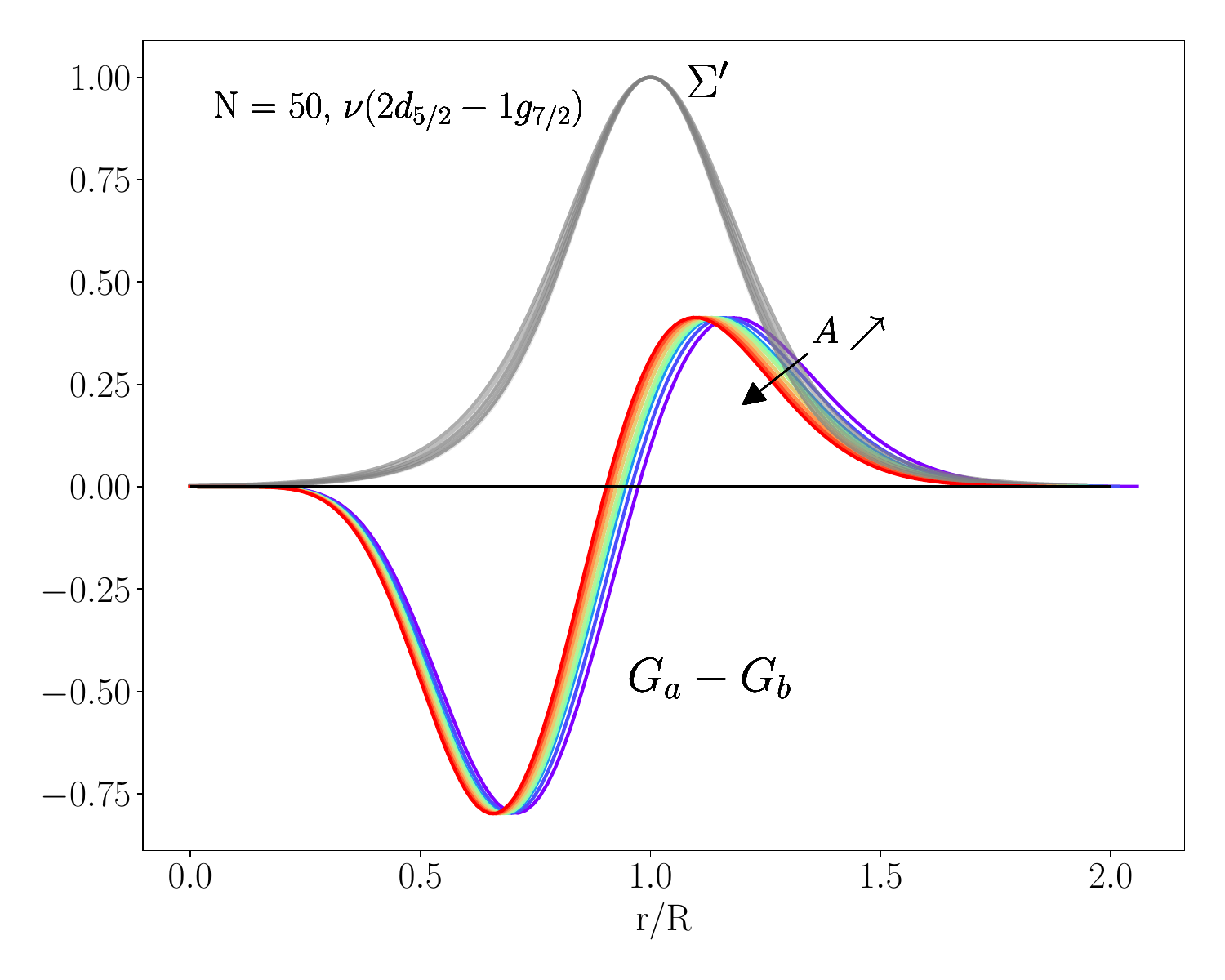}}
 \caption{Respective evolution of normalized $\Sigma'$ (dashed curves) and $Ga-G_b$ the integrated upper component densities (plain curves) along N = 50 isotonic chain, for $1\tilde{f}$ partners.}
 \label{fig:N50_G1G2}
\end{figure}

We first analyze the role of the term related to the upper components. It is possible to understand the typical scaling with $A$ by analyzing the typical behaviours of the two integrands of $\Delta E_g$ (see Eq. (\ref{Eq:PT1PSS})), displayed on Fig. \ref{fig:N50_G1G2} for increasing value of $A$. In the case of $SS$, $F_\downarrow - F_\uparrow$ was always positive and featured a clear peak. In the present case, $G_a - G_b$ has a more involved structure, as $a$ and $b$ have different $n$ and $\ell$. Nevertheless, $G_a-G_b$ is found to be negative in the core of the nucleus and changes sign around its surface. This behavior is generic: $G_a-G_b$ is typically negative in the nuclear core and positive near the surface. As $A$ increases, $R$ increases and $\Sigma'$ is shifted linearly with $R$. On the other hand, as for $F_\downarrow-F_\uparrow$, $G_a-G_b$ is shifted typically as the harmonic oscillator length (since they are eigenvectors of a SS-RHO hamiltonian) that scales as $R^{1/2}$. Therefore, as $A$ increases, $\Sigma'$ probes more and more the positive part of $G_a-G_b$. This leads to an increase of $\Delta E_g$ with $A$, as $\Sigma'$ increasingly probes the positive lobe of $G_a-G_b$. Two points should be considered: i) The typical evolution with $A$ is difficult to capture, as it is very sensitive to the details of both $\Sigma'$ and $G_a - G_b$. ii) As a rule of thumb, we found that the general scaling is roughly linear with $A$ for PSS partners, on top of a magic gap formed by SS partners, explaining the behavior of $\Delta E_g$ observed on Fig. \ref{fig:N50_EgEf}.

The present analysis allows to understand the role of $\tilde{\ell}$: as $\tilde{\ell}$ increases, $G_1$ and $G_2$ are shifted away from the center of the nucleus since their angular momentum also increases. This ensures a decrease of $\Delta E_g$, as the overlap between $\Sigma'$ and $G_2-G_1$ diminishes.

\subsubsection{$\Delta E_f$}

\begin{figure}[tb]
\scalebox{0.35}{\includegraphics{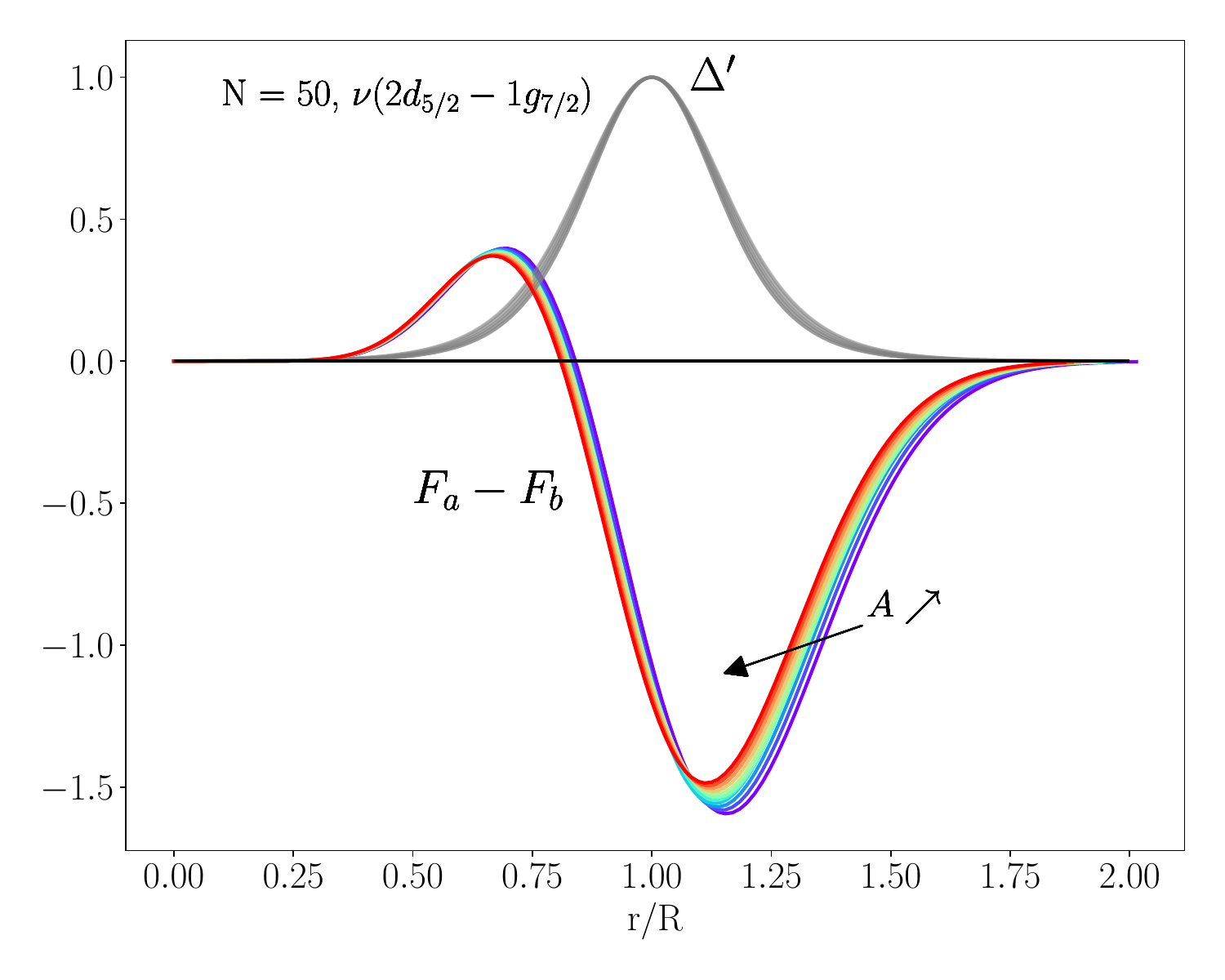}}
 \caption{Respective evolution of normalised $\Delta'$ (dashed curves) and $F_a-F_b$ the integrated lower component densities (plain curves) along N = 50 isotonic chain, for $1\tilde{f}$ partners.}
 \label{fig:N50_F1F2}
\end{figure}

We now focus on the second term of the r.h.s. of Eq. (\ref{Eq:PT1PSS}). In the case of $\Delta E_f$,  the lower components wavefunctions $f_a$ and $f_b$ have similar structures, leading to a smoother behavior: in the limit case of exact PSS, $f_a = -f_b$ \cite{ginocchio2002,ginocchio2004}. Both integrands of $\Delta E_f$ are depicted on Fig. \ref{fig:N50_F1F2}.

There are two competing  effects leading to an approximate constant value of $\Delta E_f$ as a function of $A$, observed on Fig. \ref{fig:N50_EgEf}. The first is kinematical and tends to decrease $\Delta E_f$. This is similar to the SS case : as $A$ increases, $\omega$ decreases and therefore the norm of $F_{a,b}$ decreases, according to equation (\ref{Eq:normf}). The second effect is geometrical and tends to increase the value of $\Delta E_f$: as $A$ increases, the overlap between the positive lobe of $F_a-F_b$ and  $\Delta'$ increases, yielding a larger value. In the end, these two effects cancel out and yield an almost constant contribution as a function of $A$.

When $\tilde{\ell}$ increases, not only is $F_a-F_b$  shifted towards the surface, its height also increases, due once again to Eq. (\ref{Eq:normf}). Here, there are two competitive features at stake: i) As its height increases, it increases $\Delta E_f$ in absolute value, and ii) If $\tilde{\ell}$ is too large, $|\Delta E_f|$ decreases as $\Delta'$ hits the positive lobe and the negative lobe is no longer probed.

\subsubsection{General mechanisms to PSS breaking}

In light of the present analysis, we offer a systematic framework to better understand PSS splitting, grounded in a perturbative approach. It should be noted that a previous study \cite{alberto2002} made a significant progress in exploring the mechanisms of PSS breaking, though it relied on a non-perturbative method where energy splittings involved divergent integrals, requiring regularization. In contrast, the present perturbative treatment provides closed-form expressions that simplify the interpretation of splitting mechanisms by reducing them to at most two competing contributions. This not only enhances the clarity of the results but also unifies the treatment of both SS and PSS splittings, offering a broader perspective.

As the contribution of $\Delta $ always yields an almost constant negative value, its precise shape is not relevant to the present discussion. This can be traced back to the radial structure of both $\Delta'$ and $F_a-F_b$ and is consistent with the findings of \cite{alberto2002}. 

On the other hand, the deeper the depth of $\Sigma$, the larger the absolute PSS splitting, as can be seen from Eq. (\ref{Eq:PT1PSS}). This is natural, the commutator of PSS generator with the Dirac Hamiltonian being proportional to $\Sigma'$. In addition, the role of the  diffusivity is subtle. In  \cite{alberto2002}, it was found that the splitting decreases, and eventually becomes negative, as the diffusivity increases. This can now be easily understood with Fig. \ref{fig:N50_G1G2}: as the diffusivity increases, $\Sigma'$ gets wider, probing more and more the negative $G_a-G_b$ lobe.

Finally, it should be noted that the PSS splitting can be decomposed as follows: 

\begin{align}
    \Delta E_\textrm{PSS} &= E_{n,\ell,\ell+1/2} - E_{n-1,\ell+2,\ell+3/2} \\
    & = E_{n,\ell,\ell+1/2}-E_{n-1,\ell+2,\ell+1/2} \nonumber \\
    & \textrm{  }-E_{n-1,\ell+2,\ell+3/2}+E_{n-1,\ell+2,\ell+1/2} \\
    & = \Delta E_\textrm{HO}(N,\ell+1/2) - \Delta E_\textrm{SO}(n-1,\ell+2)
\label{EQ:deco}    
\end{align}

\noindent The first term, $\Delta E_\textrm{HO}$, comes from the central potential, breaking the $N = 2(n-1)+\ell$ accidental degeneracy of the HO. The second one, $\Delta E_\textrm{SO}$, comes from spin-orbit splitting of orbitals $(n-1,\ell'=\ell+2, j = \ell' \pm 1/2)$. $\Delta E_\textrm{HO}$ can be identified with $\Delta E_g$, and $\Delta E_\textrm{SO}$ with $-\Delta E_f$. The spin symmetry being always negatively broken (spin down being lower in energy than spin up), it ensures that $\Delta E_{\textrm{SO}}<0$.  On the other hand, $\Delta E_{\textrm{HO}}$ may yields positive or negative values, depending on the radial structure of the orbitals at stake. Eq. (\ref{EQ:deco}) enlightens the contributions of both $\Sigma$ and $\Delta$ in PSS splitting, whereas only $\Delta$ was involved in SS breaking.

\section{Conclusion}

Spin and pseudo-spin symmetries breakings in relativistic mean-field approaches are studied within a common perturbative framework. 
We derive for the first time an explicit closed formula for both spin and pseudo-spin gaps. For spin symmetry, it is found that only the $\Delta = S-V$ and the lower component $f$ are involved. Within this framework, two effects are at stake for generic SS splitting scaling with $A$. i) The decrease of the norm of $f$ due to the relativistic nature of $\Psi$ ii) As $A$ increases, $\Delta'$ probes the decreasing part of $F_\downarrow-F_\uparrow$, leading to another attenuating factor. In the case of PSS splitting, both $\Sigma$ and $\Delta$ potential and both $f$ and $g$ components are involved. The global trend with $A$ is here driven by $g$ and $\Sigma$, whereas $\Delta$ and $f$ yield an almost constant negative value as a function of mass number. The dependence on the precise shape of the potentials have also been discussed.

With this work, spin and pseudo-spin symmetries can be treated on the same footing and allows to disentangle and clarify the roles of different nuclear potential in their breaking or restoration. Typical trends in pseudo-spin splittings can be captured at first order in perturbation theory. However, a more detailed understanding of the accidental restoration of pseudo-spin symmetry in specific nuclei—underlying several nuclear phenomena—requires a deeper analysis.  In particular, second-order perturbation theory offers a refined description by more accurately determining the absolute value of the splittings, thereby providing deeper insight into the mechanisms governing pseudo-spin symmetry realization.

It would also be interesting to probe PSS splitting in neutron-rich nuclei. They are known to develop a neutron skin, resulting in a particularly large diffusivity \cite{dobaczewski1994}. Such nuclei could be excellent benchmarks of the effect of the diffusivity on PSS splitting. It would also be interesting to look for such a splitting in bubble nuclei, which have a depleted density in their core. This would result in a negative value of $\Sigma'$ in the center of the nucleus, which in turn would increase PSS splitting, $G_a-G_b$ being negative in the core of the nucleus. 


\newpage
\appendix

\bigskip

\bibliography{pss_bib}



\end{document}